\documentclass[11pt]{article}
\usepackage{amssymb,amsmath}
\usepackage{graphicx}
\usepackage{epsfig}
\begin{document}

\title{Entangled state representations in noncommutative quantum
mechanics
\thanks{This project supported by National Natural
Science Foundation of China under Grant 10375056 and 90203002.}}
\author{{Sicong Jing $^{1}$, Qiu-Yu Liu $^{1}$ and Hongyi Fan $^{2}$}\\
$^{1}${\small Department of Modern Physics, University of Science and Technology of China,}\\
{\small Hefei, Anhui 230026, China}\\
$^{2}${\small Department of Material Science and Engineer,
University of Science and Technology of China,}\\
{\small Hefei, Anhui 230026, China}}
\maketitle

\begin{abstract}
We introduce new representations to formulate quantum mechanics on
noncommutative coordinate space, which explicitly display
entanglement properties between degrees of freedom of different
coordinate components and hence could be called entangled state
representations. Furthermore, we derive unitary transformations
between the new representations and the ordinary one used in
noncommutative quantum mechanics (NCQM) and obtain eigenfunctions
of some basic operators in these representations. To show the
potential applications of the entangled state representations, a
two-dimensional harmonic oscillator on the noncommutative plane
with both coordinate-coordinate and momentum-momentum couplings is
exactly solved. \\


\end{abstract}

\section{Introduction}
As is well known, representations and transformation theories,
founded by Dirac \cite{s1}, play basic and important role in
quantum mechanics. Many quantum mechanics problems were solved
cleverly by working in specific representations. Some
representations, such as, the coordinate, the momentum, the number
representation, as well as the coherent state representation, are
often employed in the literature of ordinary quantum mechanics. In
noncommutative quantum mechanics (NCQM) \cite{s2}, because of the
non-commutativity of coordinate-component operators, there are no
common eigenstates for these different coordinate operators, and
one can hardly construct a coordinate representation in the usual
sense. However, in order to formulate quantum mechanics on a
noncommutative space so that some dynamic problems can be solved,
we do need some appropriate representations. On the other hand, we
realize that in NCQM ordinary products are usually replaced by
$*$-products between functions on the noncommutative space
\cite{s3}, which is equivalent to work in some kind of
"quasi-coordinate" representation (in this representation, the
state vectors, for example, $|x,y>$, are not common eigenstates of
the coordinate operators $\hat{X}$ and $\hat{Y}$ in NCQM. For
details, please see Sec. 4). Of course, if we have more practical
representations for NCQM, it will be more powerful to deal with
the dynamic problems in NCQM.

\par
Noticing that although two coordinate-component operators on the
noncommutative space do not commute each other, the difference of
the two coordinate operators indeed commute with the sum of the
relevant two momentum operators, thus we can still employ
Einstein-Podolsky-Rosen's (EPR) \cite{s4} idea to construct
entangled states on the noncommutative space. It is easily to show
that the entangled states with continuum variables are orthonormal
and satisfy completeness relations, therefore they present new
representations for NCQM. The first bipartite entangled state
representation of continuum variables is constructed by one of the
authors (H. Fan) and J. R. Klauder \cite{s5} based on the idea of
quantum entanglement initiated by EPR who used commutative
property of two particles' relative coordinate and total momentum.
Here we extend the formalism in \cite{s5} to NCQM and investigate
some basic properties of the entangled state representations on
the noncommutative space. We also derive explicit unitary
operators which connect the entangled state representations and
the "quasi-coordinate" representation and transfer them each
other. Using the unitary transformation, it is convenient to
obtain eigenfunctions of some basic operators of NCQM in one
representation if one knows them in another representation. To
show the potential applications of the entangled state
representations in NCQM, we solve exactly the energy level of a
two-dimensional harmonic oscillator on a noncommutative plane with
both kinetic coupling and elastic coupling.

\par
The work is arranged as follows: In Sec. 2 we construct the
entangled state representations for NCQM and derive matrix
elements of coordinate and momentum operators in these
representations. In order to demonstrate these states are indeed
the entangled states, we study their Schmidt decompositions in
Sec. 3. In Sec. 4 we investigate the transformation between the
"quasi-coordinate" and the entangled state representations, and
derive an explicit unitary operator which transforms them each
other. Sec. 5 is devoted to study eigenfunctions of some basic
operators on the noncommutative plane in the entangled state
representation, which all display some extent of entanglement
between the coordinates and the momenta. In Sec. 6 we study a
two-dimensional harmonic oscillator on a noncommutative plane with
both kinetic coupling and elastic coupling and solve its energy
spectrum exactly. Some summary and discussion are presented in
Sec. 7.

\section{Entangled state representations for NCQM}
Without loss of generality and for the sake of simplicity, we only
discuss the noncommutative plane case in the follows. Operators
$\hat{X}$, $\hat{Y}$, $\hat{P}_{x}$ and $\hat{P}_{y}$ satisfy the
following commutation relations
\begin{equation}
[\hat{X},\,\hat{Y}]=\emph{i}\,\theta,~~~~~~~~
[\hat{X},\,\hat{P}_{x}]=\emph{i},~~~~~~~~ [\hat{Y},\,\hat{P}_{y}
]=\emph{i},
\end{equation}
and other commutators of these operators are vanishing, where
$\theta$ is a real parameter reflecting the non-commutativity of
space coordinates, and we take $\hbar=1$. Considering the
following operators
\begin{equation}
\hat{R}=\frac{\hat{X}-\hat{Y}}{\sqrt{2}},~~~~
\hat{P}=\frac{\hat{P}_{x}+\hat{P}_{y}}{\sqrt{2}},~~~~
\hat{S}=\frac{\hat{X}+\hat{Y}}{\sqrt{2}},~~~~
\hat{K}=\frac{\hat{P}_{x}-\hat{P}_{y}}{\sqrt{2}}.
\end{equation}
Obviously $\hat{R}$ and $\hat{P}$ are commute each other, as well
as $\hat{S}$ and $\hat{K}$ are commute each other, respectively.
Thus $\hat{R}$ and $\hat{P}$ have common eigenstates $|\eta >$,
and $\hat{S}$ and $\hat{K}$ have common eigenstates $|\xi >$. Here
$\eta$ and $\xi$ may be complex numbers,
($\eta=\eta_{1}+\emph{i}\,\eta_{2}$, and $\xi =
\xi_{1}+\emph{i}\,\xi_{2}$), and $\eta_{1}$, $\eta_{2}$, $\xi_{1}$
and $\xi_{2}$ are real numbers.

\par
In order to get explicit expressions of the eigenstates $|\eta >$
and $|\xi >$, we use the following transformations
\begin{equation}
\hat{X}=x-\frac{\theta}{2}\,p_{y},~~~~\hat{Y}=y+\frac{\theta}{2}\,p_{x},~~~~
\hat{P}_{x}=p_{x},~~~~\hat{P}_{y}=p_{y},
\end{equation}
where the operators $x$, $y$, $p_{x}$ and $p_{y}$ satisfy ordinary
Heisenberg commutation relations
\begin{equation}
[x,\,p_{x} ]=\emph{i},~~~~[y,\,p_{y} ]=\emph{i},
\end{equation}
and other commutators of these operators are vanishing.
Furthermore, introducing two independent ordinary bosonic creation
and annihilation operators $a^{\dag}$, $a$ and $b^{\dag}$, $b$
with commutation relations $[a,\,a^{\dag} ]=1$,
$[b,\,b^{\dag}]=1$, we have
\begin{equation}
x=\frac{a+a^{\dag}}{\sqrt{2}},~~~~
p_{x}=\frac{a-a^{\dag}}{\sqrt{2}\,\emph{i}},~~~~
y=\frac{b+b^{\dag}}{\sqrt{2}},~~~~
p_{y}=\frac{b-b^{\dag}}{\sqrt{2}\,\emph{i}}.
\end{equation}
In terms of these creation and annihilation operators, we can
express the operators $\hat{X}$, $\hat{Y}$, $\hat{P}_{x}$ and
$\hat{P}_{y}$ as
\begin{eqnarray}
&&\hat{X}=\frac{a+a^{\dag}}{\sqrt{2}} -\frac{\theta
(b-b^{\dag})}{2\sqrt{2}\emph{i}},~~~~
\hat{P}_{x}=\frac{a-a^{\dag}}{\sqrt{2}\,\emph{i}},\nonumber\\
&&\hat{Y}=\frac{b+b^{\dag}}{\sqrt{2}} +\frac{\theta
(a-a^{\dag})}{2\sqrt{2}\emph{i}},~~~~
\hat{P}_{y}=\frac{b-b^{\dag}}{\sqrt{2}\,\emph{i}}.
\end{eqnarray}
Thus the operators $\hat{R}$ and $\hat{P}$ may be expressed as
\begin{equation}
\hat{R}=\frac{1}{2}\left(a+a^{\dag}-b-b^{\dag}\right) -
\frac{\theta}{4\emph{i}}\left(a-a^{\dag}+b-b^{\dag} \right),~~~~
\hat{P}=\frac{1}{2\emph{i}}\left(a-a^{\dag}+b-b^{\dag}\right).
\end{equation}

\par
The common eigenstate $|\eta >$ of $\hat{R}$ and $\hat{P}$ can be
written as
\begin{equation}
|\eta >=\frac{1}{\sqrt{\pi}}\exp{\left(-\frac{|\eta|^{2}}{2}+\eta
a^{\dag}-\eta^{\ast} b^{\dag}+ a^{\dag} b^{\dag}\right)} |00>,
\end{equation}
where $|00>$ is a two-mode bosonic vacuum state satisfying
$a\,|00>=0$ and $b\,|00>=0$. It is easily to see that
\begin{equation}
\frac{1}{2}\left(a+a^{\dag}-b-b^{\dag}\right)
|\eta>=\eta_{1}|\eta>,~~~~
\frac{1}{2\emph{i}}\left(a-a^{\dag}+b-b^{\dag}\right)
|\eta>=\eta_{2}|\eta>,
\end{equation}
which lead to
\begin{equation}
\hat{R}\,|\eta>=\left(\eta_{1}-\frac{\theta}{2}\eta_{2}\right)|\eta>,~~~~
\hat{P}\,|\eta>=\eta_{2}|\eta >.
\end{equation}
Here we would like to give an explicitly proof of the completeness
relation for the eigenstates $|\eta >$ using a method of
Integration within Ordered Product (IWOP) of products \cite{s6}
\begin{eqnarray}
&&\int_{-\infty}^{\infty}d^{2}\eta\,|\eta>\,<\eta|=
\int_{-\infty}^{\infty}\frac{d^{2}\eta}{\pi}\,:\exp{(-|\eta|^{2}+
\eta a^{\dag}-\eta^{\ast}b^{\dag}+a^{\dag}b^{\dag}-a^{\dag}a-
b^{\dag}b+\eta^{\ast}a-\eta b +a b)}:\nonumber\\
&&=:\exp{\left((a^{\dag}-b)(a-b^{\dag})+a^{\dag}b^{\dag}+a
b-a^{\dag}a-b^{\dag}b\right)}:=1,
\end{eqnarray}
where $d^{2}\eta \equiv d\eta_{1}d\eta_{2}$ and we have used an
expression $|00>\,<00|=:\exp{(-a^{\dag}a-b^{\dag}b)}:$ and the
notation $:...:$ stands for taking the normal product of the
creation and annihilation operators. It is easily to derive the
inner product of the states $|\eta >$
\begin{equation}
<\eta|\eta '>=\delta^{(2)} (\eta -\eta ')=\delta
(\eta_{1}-\eta_{1}')\delta(\eta_{2}-\eta_{2}').
\end{equation}
Therefore, the eigenstates $|\eta >$ form an orthonormal and
complete set of base vectors and can be used to expand any other
state-vector in the related Hilbert space, so these state form a
representation for NCQM.

\par
Similarly, we may express the operators $\hat{S}$ and $\hat{K}$ as
\begin{equation}
\hat{S}=\frac{1}{2}(a+a^{\dag}+b+b^{\dag})+\frac{\theta}{4\emph{i}}
(a-a^{\dag}-b+b^{\dag}),~~~~
\hat{K}=\frac{1}{2\emph{i}}(a-a^{\dag}-b+b^{\dag}).
\end{equation}
The common eigenstate of $\hat{S}$ and $\hat{K}$ is
\begin{equation}
|\xi >=\frac{1}{\sqrt{\pi}}\exp{\left(-\frac{|\xi|^{2}}{2}+\xi
a^{\dag}+ \xi^{\ast}b^{\dag}-a^{\dag}b^{\dag}\right)}|00>.
\end{equation}
With the aid of two expressions
\begin{equation}
\frac{1}{2}(a+a^{\dag}+b+b^{\dag})|\xi >=\xi_{1}|\xi >,~~~~
\frac{1}{2\emph{i}}(a-a^{\dag}-b+b^{\dag})|\xi >=\xi_{2}|\xi >,
\end{equation}
we have
\begin{equation}
\hat{S}\,|\xi >=\left( \xi_{1} +
\frac{\theta}{2}\xi_{2}\right)|\xi
>, ~~~~\hat{K}\,|\xi >=\xi_{2}|\xi >.
\end{equation}
Also the states $|\xi >$ form an orthonormal and complete set of
base vectors
\begin{equation}
\int_{-\infty}^{\infty}d^{2}\xi\,|\xi >\,<\xi|=1, ~~~~
<\xi|\xi'>=\delta^{(2)}(\xi-\xi')=\delta(\xi_{1}-\xi_{1}')\delta(\xi_{2}-\xi_{2}'),
\end{equation}
here $d^{2}\xi \equiv d\xi_{1}d\xi_{2}$.

\par
Thus the eigenstates $|\eta >$ and $|\xi >$ form two
representations for quantum mechanics on the noncommutative plane,
respectively. In the next section we will explain that in fact the
states $|\eta >$ and $|\xi >$ basically are entangled states in
the noncommutative plane, so we may call the $|\eta >$ and $|\xi
>$ representations as entangled state representations. For the
noncommutative quantum plane, sometimes working in the $|\eta
>$ or $|\xi >$ representation is more convenient, so we first need to
know the scalar product of $|\eta >$ and $|\xi >$. With the aid of
over-completeness of coherent states
\begin{equation}
\int \frac{d^{2}z_{1}d^{2}z_{2}}{\pi^{2}}|z_{1},z_{2}>
\,<z_{1},z_{2}|=1,
\end{equation}
where $|z_{1},z_{2}>$ is a two-mode canonical coherent state
\begin{equation}
|z_{1},z_{2}>=|z_{1}>_{a}|z_{2}>_{b}=e^{-\frac{1}{2}(|z_{1}|^{2}+|z_{2}|^{2})}
e^{z_{1}a^{\dag}+z_{2}b^{\dag}}|00>,
\end{equation}
one may simply get
\begin{equation}
<\eta|\xi>=\int \frac{d^{2}z_{1}d^{2}z_{2}}{\pi^{2}}<\eta|
z_{1},z_{2}> \,<z_{1},z_{2}|\xi>=\frac{1}{2\pi}
e^{\emph{i}(\eta_{1}\xi_{2}-\eta_{2}\xi_{1})}.
\end{equation}
Having the eq.(20), one easily gets all of matrix elements of the
basic operators $\hat{X}$, $\hat{Y}$, $\hat{P}_{x}$ and
$\hat{P}_{y}$ on the noncommutative plane in the entangled state
representation $|\eta >$. To do this, we only need to evaluate
$<\eta|\hat{S}|\eta'>$ and $<\eta|\hat{K}|\eta'>$, and obtain
\begin{equation}
<\eta|\hat{S}|\eta'>=<\eta|\hat{S}\int d^{2}\xi
|\xi>\,<\xi|\eta'>=\emph{i}\,\left(\frac{\partial}{\partial
\eta_{2}}-\frac{\theta}{2}\frac{\partial}{\partial
\eta_{1}}\right)\delta^{(2)}(\eta -\eta'),
\end{equation}
and
\begin{equation}
<\eta|\hat{K}|\eta'>=<\eta|\hat{K}\int d^{2}\xi
|\xi>\,<\xi|\eta'>=-\emph{i}\,\frac{\partial}{\partial
\eta_{1}}\delta^{(2)}(\eta -\eta').
\end{equation}
Thus in the $|\eta >$ representation, we have
\begin{eqnarray}
&&<\eta|\hat{X}|\eta'>=\frac{1}{\sqrt{2}}\left(\eta_{1}
+\emph{i}\,\partial_{\eta_{2}}-\frac{\theta}{2}(\eta_{2}
+\emph{i}\,\partial_{\eta_{1}})\right)\delta^{(2)}(\eta
-\eta'),\nonumber\\
&&<\eta|\hat{Y}|\eta'>=\frac{-1}{\sqrt{2}}\left(\eta_{1}
-\emph{i}\,\partial_{\eta_{2}}-\frac{\theta}{2}(\eta_{2}
-\emph{i}\,\partial_{\eta_{1}})\right)\delta^{(2)}(\eta
-\eta'),\nonumber\\
&&<\eta|\hat{P}_{x}|\eta'>=\frac{1}{\sqrt{2}}\left(\eta_{2}
-\emph{i}\,\partial_{\eta_{1}}\right)\delta^{(2)}(\eta
-\eta'),\nonumber\\
&&<\eta|\hat{P}_{y}|\eta'>=\frac{1}{\sqrt{2}}\left(\eta_{2}
+\emph{i}\,\partial_{\eta_{1}}\right)\delta^{(2)}(\eta -\eta').
\end{eqnarray}

\par
Similarly, in the $|\xi >$ representation, we have
\begin{eqnarray}
&&<\xi|\hat{X}|\xi'>=\frac{1}{\sqrt{2}}\left(\xi_{1}
+\emph{i}\,\partial_{\xi_{2}}+\frac{\theta}{2}(\xi_{2}
+\emph{i}\,\partial_{\xi_{1}})\right)\delta^{(2)}(\xi
-\xi'),\nonumber\\
&&<\xi|\hat{Y}|\xi'>=\frac{1}{\sqrt{2}}\left(\xi_{1}
-\emph{i}\,\partial_{\xi_{2}}+\frac{\theta}{2}(\xi_{2}
-\emph{i}\,\partial_{\xi_{1}})\right)\delta^{(2)}(\xi
-\xi'),\nonumber\\
&&<\xi|\hat{P}_{x}|\xi'>=\frac{1}{\sqrt{2}}\left(\xi_{2}
-\emph{i}\,\partial_{\xi_{1}}\right)\delta^{(2)}(\xi
-\xi'),\nonumber\\
&&<\xi|\hat{P}_{y}|\xi'>=\frac{-1}{\sqrt{2}}\left(\xi_{2}
+\emph{i}\,\partial_{\xi_{1}}\right)\delta^{(2)}(\xi -\xi').
\end{eqnarray}

\section{Entanglement properties of the states $|\eta >$ and $|\xi >$}
From eq.(8) and eq.(14) we find that there exists intrinsic
entanglement of different degrees of freedom corresponding to
different coordinate components on a noncommutative plane.
Usually, these states are so-called entangled states, therefore we
may name these two representations as entangled state
representations. In order to show this kind of entanglement more
explicitly, let us consider Fourier transform of the state $|\eta
>$. Using a familiar expression for eigenstate $|q>$ of coordinate
operator $x$ ($x|q>=q|q>$) in Fock space
\begin{equation}
|q>_{a}=\frac{1}{\sqrt[4]{\pi}} \exp{ \left(
-\frac{q^{2}}{2}+\sqrt{2}q a^{\dag}-\frac{a^{\dag
2}}{2}\right)}|0>,
\end{equation}
one can write the Fourier transform of $|\eta >$ as
\begin{equation}
\int_{-\infty}^{\infty}\frac{d\eta_{2}}{\sqrt{2\pi}}\,|\eta >\,
e^{-\emph{i}u\eta_{2}}=\left|\frac{u+\eta_{1}}{\sqrt{2}}\right>_{a}\,
\left|\frac{u-\eta_{1}}{\sqrt{2}}\right>_{b}.
\end{equation}
If one furthermore consider inverse Fourier transform of the above
expression, one will get
\begin{equation}
|\eta >=\frac{1}{\sqrt{\pi}}e^{-\emph{i}\eta_{1}\eta_{2}}
\int_{-\infty}^{\infty}\,dq\,\Big|q\Big>_{a}\,
\left|q-\sqrt{2}\eta_{1}\right>_{b}e^{\emph{i}\sqrt{2}\eta_{2}q}.
\end{equation}
This is exactly the well-known Schmidt decomposition of a pure
state which expresses the pure state can not be factorized as a
direct product of two other states and therefore is an entangled
state. On the other hand, noticing expression for eigenstate $|p>$
of momentum operator $p$ in the Fock space
\begin{equation}
|p>_{a}=\frac{1}{\sqrt[4]{\pi}} \exp{ \left(
-\frac{p^{2}}{2}+\emph{i}\sqrt{2}p a^{\dag}+\frac{a^{\dag
2}}{2}\right)}|0>,
\end{equation}
one can also derive
\begin{equation}
\int_{-\infty}^{\infty}\frac{d\eta_{1}}{\sqrt{2\pi}}\,|\eta >\,
e^{\emph{i}v\eta_{1}}=\left|\frac{v+\eta_{2}}{\sqrt{2}}\right>_{a}\,
\left|\frac{-v+\eta_{2}}{\sqrt{2}}\right>_{b}
\end{equation}
in terms of the eigenstates of the momentum operator whose inverse
Fourier transform leads to another standard expression for an
entangled state
\begin{equation}
|\eta >=\frac{1}{\sqrt{\pi}}e^{-\emph{i}\eta_{1}\eta_{2}}
\int_{-\infty}^{\infty}\,dp\,\left|p+\sqrt{2}\eta_{2}\right>_{a}\,
\Big|-p\Big>_{b}e^{-\emph{i}\sqrt{2}\eta_{1}p}.
\end{equation}

\par
For the eigenstate $|\xi >$, using the eigenstate $|q>$ of
coordinate operator (eq.(25)), one has similarly
\begin{equation}
\int_{-\infty}^{\infty}\frac{d\xi_{2}}{\sqrt{2\pi}}\,|\xi >\,
e^{-\emph{i}u\xi_{2}}=\left|\frac{u+\xi_{1}}{\sqrt{2}}\right>_{a}\,
\left|\frac{u-\xi_{1}}{\sqrt{2}}\right>_{b}.
\end{equation}
Its inverse Fourier transform is the Schmidt decomposition of the
state $|\xi >$
\begin{equation}
|\xi >=\frac{1}{\sqrt{\pi}}e^{\emph{i}\xi_{1}\xi_{2}}
\int_{-\infty}^{\infty}\,dq\,\left|q+\sqrt{2}\xi_{1}\right>_{a}\,
\Big|-q\Big>_{b}e^{\emph{i}\sqrt{2}\xi_{2}q}.
\end{equation}
Of course, in terms of the eigenstate $|p>$ of the momentum
operator (eq.(28)), one can get
\begin{equation}
\int_{-\infty}^{\infty}\frac{d\xi_{1}}{\sqrt{2\pi}}\,|\xi >\,
e^{\emph{i}v\xi_{1}}=\left|\frac{v+\xi_{2}}{\sqrt{2}}\right>_{a}\,
\left|\frac{v-\xi_{2}}{\sqrt{2}}\right>_{b},
\end{equation}
and its inverse transform leads to another Schmidt decomposition
of the state $|\xi >$
\begin{equation}
|\xi >=\frac{1}{\sqrt{\pi}}e^{\emph{i}\xi_{1}\xi_{2}}
\int_{-\infty}^{\infty}\,dp\,\Big|p\Big>_{a}\,
\left|p-\sqrt{2}\xi_{2}\right>_{b}e^{-\emph{i}\sqrt{2}\xi_{1}p}.
\end{equation}
Therefore it is reasonable to call the $|\eta >$ and $|\xi>$
representations as entangled state representations.

\section{Unitary transformations}
In the past years NCQM was discussed extensively in various
aspect. The most popular method of formulating NCQM in the vast
literature is treating the coordinates as commuting, but
introducing $*_{\theta}$-product (for instance, in the
noncommutative plane, $*_{\theta}\equiv
\exp{\frac{\emph{i}\,\theta}{2}\left(\overleftarrow{\partial}_{x}\,
\overrightarrow{\partial}_{y}-\overleftarrow{\partial}_{y}\,
\overrightarrow{\partial}_{x}\right)}$) between functions on the
noncommutative space to reflect the non-commutativity of
coordinates. Using the $*_{\theta}$-product, Schr\"{o}dinger
equation \begin{equation}
\hat{H}(\hat{X},\hat{Y},\hat{P}_{x},\hat{P}_{y})|\psi>=E|\psi>
\end{equation}
on the noncommutative plane should be written as \cite{s7}
\begin{equation}
\hat{H}(x,y,p_{x},p_{y})*_{\theta}\psi(x,y)=E\psi(x,y)
\end{equation}
where $\psi(x,y)=<x,y|\psi>$, the operators $\hat{X}$, $\hat{Y}$,
$\hat{P}_{x}$ and $\hat{P}_{y}$ satisfy the commutation relations
(1), and the operators $x$, $y$, $p_{x}$ and $p_{y}$ satisfy the
commutation relations (4), respectively.  It fact, in eq.(36)
people have used the representation $|x,y>=|x>_{a}\,|y>_{b}$ which
is common eigenstate of the operators $x$ and $y$ (not the
operators $\hat{X}$ and $\hat{Y}$), so we would like to name it
the "quasi-coordinate" representation. In the $|x,y>$
representation one can simply write out the matrix elements of the
operators $\hat{X}$, $\hat{Y}$, $\hat{P}_{x}$ and $\hat{P}_{y}$ on
the noncommutative plane
\begin{eqnarray}
&&<x,y|\hat{X}|x',y'>=\left(x+\frac{\emph{i}\,\theta}{2}\partial_{y}\right)
\delta(x-x')\delta(y-y'),\nonumber\\
&&<x,y|\hat{Y}|x',y'>=\left(y-\frac{\emph{i}\,\theta}{2}\partial_{x}\right)
\delta(x-x')\delta(y-y'),\nonumber\\
&&<x,y|\hat{P}_{x}|x',y'>=-\emph{i}\,\partial_{x}\,
\delta(x-x')\delta(y-y'), \nonumber\\
&&<x,y|\hat{P}_{y}|x',y'>=-\emph{i}\,\partial_{y}\,
\delta(x-x')\delta(y-y').
\end{eqnarray}
From eq.(3) we know that the states $|x,y>$ are also common
eigenstates of the operators $\hat{X}+\frac{\theta}{2}\hat{P}_{y}$
and $\hat{Y}-\frac{\theta}{2}\hat{P}_{x}$, so there is a unitary
transformation between the two representations ($|\eta>$ and
$|x,y>$) whose matrix elements may be written as
\begin{equation}
<\eta|x,y>=\frac{1}{\sqrt{\pi}}e^{\emph{i}(\eta_{1}-\sqrt{2}x)\eta_{2}}\,
\delta(x-y-\sqrt{2}\eta_{1}),
\end{equation}
where eq.(27) is used. Similarly, using eq.(32), one can get
matrix elements of the transformation between another two
representations ($|\xi>$ and $|x,y>$)
\begin{equation}
<\xi|x,y>=\frac{1}{\sqrt{\pi}}e^{-\emph{i}(\xi_{1}-\sqrt{2}y)\xi_{2}}\,
\delta(x+y-\sqrt{2}\xi_{1}).
\end{equation}
Using the completeness relations eqs.(11), (17) and
$\int\,dxdy\,|x,y>\,<x,y|=1$, one can easily see that eqs.(38) and
(39) indeed present the unitary transformations between the
entangled state representations and the "quasi-coordinate"
representation $|x,y>$.

\par
In order to get an clear form of the unitary transformation
between the $|\eta >$ and the $|x,y>$ representations, let us
consider the following integration built from the entangled state
$|\eta>$ and the two-mode "quasi-coordinate" eigenstate $|x,y>$
\begin{eqnarray}
U&=&\int d^{2}\eta\ ,|x,y>\,<\eta|\,|_{x=\frac{\eta_{1}
+\eta_{2}}{\sqrt{2}},
y=\frac{\eta_{2}- \eta_{1}}{\sqrt{2}}}\nonumber\\
&=&\int \frac{d \eta_{1} d \eta_{2}}{\pi} \exp{\left(
-\frac{x^{2}}{2}-\frac{y^{2}}{2}+\sqrt{2}x\,a^{\dag}
+\sqrt{2}y\,b^{\dag} -\frac{a^{\dag 2}}{2}- \frac{b^{\dag
2}}{2}\right)}\nonumber\\
&\,&~~~~~~~~~~~~|00>\,<00|\exp{\left(-\frac{|\eta|^{2}}{2}
+\eta^{\ast} a -\eta \,b +a\,b\right)}\Big|_{x=\frac{\eta_{1}
+\eta_{2}}{\sqrt{2}}, y=\frac{\eta_{2}- \eta_{1}}{\sqrt{2}}},
\end{eqnarray}
here we have taken all parameters in the ordinary harmonic
oscillator expressions (i.e. $m$, $\omega$) equal to $1$ for the
simplicity. Using the trick in the calculation of eq.(11), we
obtain
\begin{equation}
U=\,:\exp{\left(-\frac{1+\emph{i}}{2}(a^{\dag}a+
b^{\dag}b+a^{\dag}b+b^{\dag}a)\right)}:,
\end{equation}
where the result of the integration is expressed in terms of the
normal ordered product. To show the unitary property of the
operator $U$ more clearly, introducing an operator
$e^{\emph{i}\,r\,S}$ with $S=a^{\dag}a+
b^{\dag}b+a^{\dag}b+b^{\dag}a$, we have
$[S,\,a^{\dag}]=a^{\dag}+b^{\dag}$ and
$[S,\,b^{\dag}]=a^{\dag}+b^{\dag}$, which lead to
\begin{equation}
e^{\emph{i}\,r\,S}\,a^{\dag}\,e^{-\emph{i}\,r\,S}=
\frac{e^{2\emph{i}\,r}+1}{2}a^{\dag}+\frac{e^{2\emph{i}\,r}-1}{2}b^{\dag},~~~~
e^{\emph{i}\,r\,S}\,b^{\dag}\,e^{-\emph{i}\,r\,S}=
\frac{e^{2\emph{i}\,r}-1}{2}a^{\dag}+\frac{e^{2\emph{i}\,r}+1}{2}b^{\dag},
\end{equation}
and further
\begin{eqnarray}
e^{\emph{i}\,r\,S}&=&e^{\emph{i}\,r\,S}\,\sum_{n,m=0}^{\infty}|n,m>\,<n,m|\nonumber\\
&=&e^{\emph{i}\,r\,S}\,\sum_{n,m=0}^{\infty}\frac{a^{\dag n}
\,b^{\dag m}} {\sqrt{n!\,m!}}|00>
\,<00|\frac{a^{n}\,b^{m}}{\sqrt{n!\,m!}}=\,:\exp{\left(
-\frac{1-e^{2\emph{i}\,r}}{2}\,S \right)}:.
\end{eqnarray}
When choosing $r=-\pi /4$, we have
\begin{equation}
U=e^{-\frac{\emph{i}\,\pi}{4}\,(a^{\dag}a+
b^{\dag}b+a^{\dag}b+b^{\dag}a)},
\end{equation}
which is unitary obviously. From eq.(44), it is easily to get
\begin{equation}
Ua^{\dag}U^{\dag}=\frac{1-\emph{i}}{2}a^{\dag}-\frac{1+\emph{i}}{2}b^{\dag},~~~~
Ub^{\dag}U^{\dag}=-\frac{1+\emph{i}}{2}a^{\dag}+\frac{1-\emph{i}}{2}b^{\dag},
\end{equation}
which lead to
\begin{equation}
U|\eta>=|x,y>\,|_{x=\frac{\eta_{1} +\eta_{2}}{\sqrt{2}},
y=\frac{\eta_{2}- \eta_{1}}{\sqrt{2}}},~~~~U|x,y>=|\eta^{\ast}>\,
|_{\eta_{1}=\frac{x-y}{\sqrt{2}},\eta_{2}=\frac{x+y}{\sqrt{2}}}.
\end{equation}
Thus $U$ indeed transfer the state $|\eta>$ to the state $|x,y>$
and vice versa. In the $|\eta>$ representation one can calculate
the following matrix element of the operator $U$:
$<\eta|U|\zeta>$, where $\zeta=\zeta_{1}+\emph{i}\,\zeta_{2}$.
Using eq.(46) one has
\begin{equation}
<\eta|U|\zeta>=<\eta|x,y>|_{x=\frac{\zeta_{1}+\zeta_{2}}{\sqrt{2}},
y=\frac{\zeta_{2}-\zeta_{1}}{\sqrt{2}}}
=\frac{1}{\sqrt{\pi}}e^{\emph{i}\,(\eta_{1}-\zeta_{1}-\zeta_{2})\eta_{2}}
\delta(\sqrt{2}\zeta_{1}-\sqrt{2}\eta_{1}).
\end{equation}
If one further takes $\zeta_{1}=(x-y)/\sqrt{2}$ and
$\zeta_{2}=(x+y)/\sqrt{2}$, eq.(47) will lead to eq.(38) exactly.
This means that eq.(38) is just a matrix element of the unitary
operator $U$ in the entangled state representation. Similarly one
can find out the unitary transformation between the $|\xi>$ and
the $|x,y>$ representations and endow eq.(39) with the same
explanation.

\par
Having the unitary transformation (38), and using
\begin{equation}
<\eta|\hat{F}|\eta'>=\int \,
dxdydx'dy'\,<\eta|x,y>\,<x,y|\hat{F}|x',y'>\,<x',y'|\eta'>,
\end{equation}
or
\begin{equation}
<x,y|\hat{F}|x',y'>=\int \, d^{2}\eta
d^{2}\eta'\,<x,y|\eta>\,<\eta|\hat{F}|\eta'>\,<\eta'|x',y'>,
\end{equation}
one may get the matrix elements of any operator $\hat{F}$ in one
representation, if one knows $\hat{F}$ in another representation.
For example, taking $\hat{F}=\hat{P}_{x}$, one has
\begin{eqnarray}
<\eta|\hat{P}_{x}|\eta'>&=&\int \,\frac{dx dy dx' dy'}{\pi}\,
e^{\emph{i}(\eta_{1}-\sqrt{2}x)\eta_{2}}\,\delta(x-y-\sqrt{2}\eta_{1})
(-\emph{i}\,\partial_{x})\delta(x-x')\delta(y-y')\nonumber\\
&&~~~~~~~~~~~~~~~~~~~~e^{-\emph{i}(\eta_{1}'-\sqrt{2}x')\eta_{2}'}\,
\delta(x'-y'-\sqrt{2}\eta_{1}')\nonumber\\
&=&e^{\emph{i}\eta_{1}(\eta_{2}-\eta_{2}')}\left(
\sqrt{2}\eta_{2}-\emph{i}\,\frac{\partial}{\partial
\sqrt{2}\eta_{1}}\right)\delta(\eta_{1}-\eta_{1}')
\delta(\eta_{2}-\eta_{2}')\nonumber\\
&=&\frac{1}{\sqrt{2}}\left(\eta_{2}-\emph{i}\,\partial
_{\eta_{1}}\right)\delta^{2}(\eta-\eta')
\end{eqnarray}
which exactly coincides with eq.(23). Similarly, one also has
\begin{eqnarray}
<\eta|\hat{P}_{y}|\eta'>&=&
e^{\emph{i}\eta_{1}(\eta_{2}-\eta_{2}')}\left(
\emph{i}\frac{\partial}{\partial
\sqrt{2}\eta_{1}}\right)\delta(\eta_{1}-\eta_{1}')
\delta(\eta_{2}-\eta_{2}')\nonumber\\
&=&\frac{1}{\sqrt{2}}\left(\eta_{2}+\emph{i}\,\partial
_{\eta_{1}}\right)\delta^{2}(\eta-\eta').
\end{eqnarray}
Noticing
\begin{equation}
x\,e^{\emph{i}(\eta_{1}-\sqrt{2}x)\eta_{2}}=\frac{1}{\sqrt{2}}\left(\eta_{1}+
\emph{i}\,\partial_{\eta_{2}}\right)e^{\emph{i}(\eta_{1}-\sqrt{2}x)\eta_{2}},
\end{equation}
one can obtain other two expressions of eq.(23). Of course, with
the aid of the unitary transformation (39), from eq.(37) one may
get eq.(24).
\par
Therefore, we derive the unitary transformations which change the
$|x,y>$ representation to the $|\eta>$ (or $|\xi>$) representation
and vise versa.

\section{Eigenfunctions of some basic operators in the entangled
state representations}

From the section 2 we see that wave function of any state vector
$|\psi>$ in the entangled state representation $|\eta>$ can be
expressed as $\psi (\eta)=\psi (\eta_{1},\eta_{2})=<\eta|\psi>$.
Eq.(23) gives the representations of the operators $\hat{X}$,
$\hat{Y}$, $\hat{P}_{x}$ and $\hat{P}_{y}$ in the entangled state
representation $|\eta>$, for example, the operators $\hat{P}_{x}$
and $\hat{P}_{y}$ can be replaced by
$\frac{1}{\sqrt{2}}\left(\eta_{2}
-\emph{i}\partial_{\eta_{1}}\right)$ and
$\frac{1}{\sqrt{2}}\left(\eta_{2}
+\emph{i}\partial_{\eta_{1}}\right)$ respectively. It is well
known that eigenfunctions of the operators $\hat{P}_{x}$ and
$\hat{P}_{y}$ in ordinary quantum mechanics mat be the plane
waves, so it is interesting to see what are the eigenfunctions of
the same operators in NCQM. Since the operators $\hat{P}_{x}$ and
$\hat{P}_{y}$ are commuting each other and have common
eigenstates, we first derive their common eigenfunctions in the
entangled state representation.

\par
Noticing eqs.(10) and (22), and $[\hat{P},\hat{K}]=0$, and
denoting common eigenstate of $\hat{P}$ and $\hat{K}$ as $|\psi>$
with eigenvalues $p$ and $k$ respectively, in the entangled state
representation we have
\begin{equation}
\psi (\eta) =
\frac{1}{\sqrt{2\pi}}\delta(\eta_{2}-p)e^{\emph{i}k\eta_{1}},
\end{equation}
where $\frac{1}{\sqrt{2\pi}}$ is normalization constant. Because
the operators $\hat{P}_{x}$ and $\hat{P}_{y}$ are linear
combination of $\hat{P}$ and $\hat{K}$, if we denote the
eigenstate of $\hat{P}_{x}$ and $\hat{P}_{y}$ as
$|\psi_{p_{x},p_{y}}>$ with eigenvalues
$p_{x}=\frac{1}{\sqrt{2}}(p+k)$ and
$p_{y}=\frac{1}{\sqrt{2}}(p-k)$ respectively, we have the common
eigenfunction of $\hat{P}_{x}$ and $\hat{P}_{y}$
\begin{equation}
\psi_{p_{x},p_{y}} (\eta) = \frac{1}{\sqrt{2\pi}}\,\delta
\left(\eta_{2}-\frac{p_{x}+p_{y}}{\sqrt{2}}\right)
e^{\emph{i}(p_{x}-p_{y})\eta_{1}/\sqrt{2}}.
\end{equation}
It is clearly shown that the eigenfunctions of $\hat{P}_{x}$ and
$\hat{P}_{y}$ in $|\eta>$ representation are entanglement of
ordinary coordinate and momentum eigenfunctions. On the other
hand, in the $|x,y>$ representation, the eigenfunctions of
$\hat{P}_{x}$ and $\hat{P}_{y}$ are simply
\begin{equation}
\psi_{p_{x},p_{y}}(x,y)=<x,y|\psi_{p_{x},p_{y}}>=
\frac{1}{2\pi}e^{\emph{i}(p_{x}x+p_{y}y)}.
\end{equation}
Using the representation transformation (38), one has
\begin{eqnarray}
\psi_{p_{x},p_{y}}(\eta)&=&\int dx dy
<\eta|x,y>\,<x,y|\psi_{p_{x},p_{y}}>\nonumber\\
&=&\int \frac{dx dy}{2\pi^{3/2}}
\delta(x-y-\sqrt{2}\eta_{1})e^{\emph{i}\left((\eta_{1}-\sqrt{2}x)\eta_{2}
+p_{x}x+p_{y}y\right)}
\end{eqnarray}
which leads to the right hand side of eq.(54) exactly. It also
means that if one wants to find eigenfunction of some operator in
the $|\eta>$ representation, one can first get the eigenfunction
in the $|x,y>$ representation and then derive the eigenfunction in
the $|\eta>$ representation by the representation transformation,
and vise versa.

\par
Now let us use this method to find out eigenfunction of $\hat{X}$
in the entangled state representation $|\eta>$. In fact it is
easily understood that the eigenfunction of $\hat{X}$ is
degenerate and in order to remove the degeneracy we should
consider common eigenfunctions of the operators $\hat{X}$ and
$\hat{P}_{y}$. From eq.(3) we see that the normalized
eigenfunction of $\hat{X}$ and $\hat{P}_{y}$ in the $|x,y>$
representation can be expressed as
\begin{equation}
\psi_{X,p_{y}}(x,y)=\frac{1}{\sqrt{2\pi}}\delta(x-\tilde{x})
e^{\emph{i}\tilde{p}_{y}y},
\end{equation}
where $\tilde{p}_{y}$ is the eigenvalue of the operator
$\hat{P}_{y}$ and $\tilde{x}-\frac{\theta}{2}\tilde{p}_{y}\equiv
\tilde{X}$ the eigenvalue of $\hat{X}$ respectively. Then using
the transformation (38) we get the common eigenfunction of
$\hat{X}$ and $\hat{P}_{y}$ in the $|\eta>$ representation
\begin{equation}
\psi_{X,p_{y}}(\eta)=\frac{1}{\sqrt{2}\,\pi}
e^{\emph{i}\eta_{1}\eta_{2}}e^{-\emph{i}\sqrt{2}\left(\tilde{X}\eta_{2}
+\tilde{p}_{y}\left(\eta_{1}+\frac{\theta}{2}\eta_{2}\right)
\right)}e^{\emph{i}\tilde{p}_{y}\left(\tilde{X}+
\frac{\theta}{2}\tilde{p}_{y}\right)}.
\end{equation}
Furthermore noticing that in the $|\eta>$ representation $\hat{X}$
and $\hat{P}_{y}$ can be expressed as
\begin{equation}
\hat{X}=\frac{1}{\sqrt{2}}\left(\eta_{1} +
\emph{i}\partial_{\eta_{2}}
-\frac{\theta}{2}\eta_{2}-\emph{i}\frac{\theta}{2}\partial_{\eta_{1}}
\right)
\end{equation}
and
\begin{equation}
\hat{P}_{y}=\frac{1}{\sqrt{2}}\left(\eta_{2}+\partial_{\eta_{1}}\right)
\end{equation}
respectively, it is straightforwardly to check
$\hat{X}\psi_{X,p_{y}}(\eta)=\tilde{X}\psi_{X,p_{y}}(\eta)$ and
$\hat{P}_{y}\psi_{X,p_{y}}(\eta)=\tilde{p}_{y}\psi_{X,p_{y}}(\eta)$.
Of course, one can also directly use the expression (59) and
solves differential equation
$\hat{X}\psi(\eta)=\tilde{X}\psi(\eta)$ in the $|\eta>$
representation to get the eigenfunction $\psi(\eta)$ of $\hat{X}$.

\par
Similarly we have common eigenfunction of $\hat{Y}$ and
$\hat{P}_{x}$ in the $|\eta>$ representation
\begin{equation}
\psi_{Y,p_{x}}(\eta)=\frac{1}{\sqrt{2}\,\pi}
e^{-\emph{i}\eta_{1}\eta_{2}}e^{-\emph{i}\sqrt{2}\left(\tilde{Y}\eta_{2}
-\tilde{p}_{x}\left(\eta_{1}+\frac{\theta}{2}\eta_{2}\right)
\right)}e^{\emph{i}\tilde{p}_{x}\left(\tilde{Y}
-\frac{\theta}{2}\tilde{p}_{x}\right)},
\end{equation}
where $\tilde{p}_{x}$ and
$\tilde{y}+\frac{\theta}{2}\tilde{p}_{x}\equiv \tilde{Y}$ are the
eigenvalue of $\hat{P}_{x}$ and $\hat{Y}$ respectively.

\par
From eqs.(58) and (61) we see that the eigenfunctions of the
operators $\hat{X}$ and $\hat{Y}$ in the entangled state
representation are not simply plane waves. In fact, they also
display some kind of entanglement of the coordinates and the
momenta. Of course, one can also obtain eigenfunctions of these
operators in the $|\xi>$ representation.

\section{Some possible applications}
It is well know that representation plays a basic role in quantum
mechanics like the coordinate systems in geometry. In section 2 we
introduced the entangled state representations $|\eta>$ and
$|\xi>$, which are related to the $|x,y>$ representation by
unitary transformations as shown in section 4. In the $|\eta>$ or
$|\xi>$ representation one can also solve Schr\"{o}dinger equation
of NCQM as in the $|x,y>$ representation, and sometimes it is more
convenient working in the entangles state representation than in
the $|x,y>$ representation. To show this, let us study a
two-dimensional harmonic oscillator on the noncommutative plane
with both momentum-momentum (kinetic) coupling and
coordinate-coordinate (elastic) coupling. The Hamiltonian is
\begin{equation}
H=\frac{1}{2}\hat{P}_{x}^{2}+\frac{1}{2}\hat{P}_{y}^{2}+
\frac{1}{2}\hat{X}^{2}+\frac{1}{2}\hat{Y}^{2}+\kappa
\hat{P}_{x}\hat{P}_{y}+\frac{\lambda}{2}
\left(\hat{X}\hat{Y}+\hat{Y}\hat{X} \right),
\end{equation}
where the operators $\hat{P}_{x}$, $\hat{P}_{y}$, $\hat{X}$ and
$\hat{Y}$ satisfy the commutation relations (1). After
substituting eq.(3) into eq.(62) we get the Hamiltonian $H$ in the
$|x,y>$ representation
\begin{eqnarray}
H&=&\frac{1}{2}\left(1+\frac{\theta}{4}\right)p_{x}^{2}+
\frac{1}{2}\left(1+\frac{\theta}{4}\right)p_{y}^{2}+
\frac{1}{2}x^{2}+\frac{1}{2}y^{2}\nonumber\\
&\,&+\left(\kappa-\frac{\lambda\,\theta^{2}}{4}\right)p_{x}p_{y}
+\lambda x y -\frac{\theta}{2}(xP_{y}-yp_{x}) +
\frac{\lambda\,\theta}{2}(xp_{x}-yp_{y}),
\end{eqnarray}
which includes not only the kinetic and the elastic coupling
terms, but also the coordinate-momentum coupling terms (they are
the angular momentum term and the squeezing term, respectively).
It is not an easy task to solve its eigenequation. However, in the
$|\eta>$ representation the Hamiltonian $H$ has simpler form
\begin{eqnarray}
H&=&\frac{1}{2}\left(1+\frac{\theta^{2}}{2}-\kappa
+\frac{\lambda\,\theta^{2}}{4}\right)p_{1}^{2}
+\frac{1}{2}(1+\lambda)p_{2}^{2}-\frac{\theta}{2}
(1+\lambda)p_{1}p_{2}\nonumber\\
&\,&+\frac{1}{2}(1-\lambda)\eta_{1}^{2}
+\frac{1}{2}\left(1+\frac{\theta^{2}}{2}+\kappa
-\frac{\lambda\,\theta^{2}}{4}\right)\eta_{2}^{2}
-\frac{\theta}{2} (1-\lambda)\eta_{1}\eta_{2},
\end{eqnarray}
where $p_{i}=-\emph{i}\,\partial/\partial\,\eta_{i}$ (i=1,2). In
the Hamiltonian (64), only the kinetic and the elastic coupling
terms survive, and it is easier to be handled than the form (63).
Of course, it is needless to emphasize that the Hamiltonian (63)
and (64) are connected via a unitary transformation described in
section 4.
\par
Before diagonalizing $H$, let us introduce some notations to
rewrite (64) so that it has more familiar form
\begin{eqnarray}
&&m_{1}=\left(1+\frac{\theta^{2}}{2}-\kappa
+\frac{\lambda\,\theta^{2}}{4} \right)^{-1},
~~~~\omega_{1}=\sqrt{(1-\lambda)
\left(1+\frac{\theta^{2}}{2}-\kappa
+\frac{\lambda\,\theta^{2}}{4}\right)},~~~~\alpha=\frac{\theta}{2}
(1+\lambda),\nonumber\\
&&m_{2}=(1+\lambda)^{-1},~~~~\omega_{2}=\sqrt{(1+\lambda)
\left(1+\frac{\theta^{2}}{2}+\kappa
-\frac{\lambda\,\theta^{2}}{4}\right)},~~~~\beta=\frac{\theta}{2}
(1-\lambda).
\end{eqnarray}
In terms of these notations, $H$ becomes
\begin{equation}
H=\frac{p_{1}^{2}}{2m_{1}}+\frac{p_{2}^{2}}{2m_{2}}-\alpha
p_{1}p_{2} + \frac{m_{1}\omega_{1}^{2}}{2}\eta_{1}^{2}+
\frac{m_{2}\omega_{2}^{2}}{2}\eta_{2}^{2} -\beta \eta_{1}\eta_{2}.
\end{equation}
Now let us introduce a two by two matrix $A$ whose matrix elements
$a_{ij}$ will be determined later ($i,j=1,2$). If we use $\vec{p}$
to denote the two-dimensional momentum $(p_{1},p_{2})$, one can
write $\vec{\tilde{p}}=A\vec{p}=(\tilde{p}_{1},\tilde{p}_{2})$
with $\tilde{p}_{i}=a_{ij}p_{j}$, and inversely,
$p_{i}=b_{ij}\tilde{p}_{j}$, where $b_{ij}$ are elements of
inverse matrix of $A$. Consider the following transformation
\begin{equation}
V=\sqrt{det\,A}\int d\vec{p}\,\,|A\vec{p}>\,<\vec{p}|
\end{equation}
in the Hilbert space spanned by two-mode momentum eigenstates
$|\vec{p}>$, which is unitary clearly
\begin{eqnarray}
V\,V^{\dag}&=&det\,A\,\int d\vec{p}\,d\vec{p}\prime|A\vec{p}>\,\,
<\vec{p}|\vec{p}\prime >\,<A\vec{p}\prime|\nonumber\\
&=&det\,A\,\int d\vec{p}\,\,|A\vec{p}>\,<A\vec{p}|=\int
d\vec{\tilde{p}}\,\,|\vec{\tilde{p}}>\,<\vec{\tilde{p}}|=1,
\end{eqnarray}
and similarly $V^{\dag}\,V=1$. In eq.(67),
$|\vec{p}>=|p_{1}>|p_{2}>$ and $|p_{i}>$ are the momentum
eigenstates
\begin{equation}
|p_{i}>=\left(\frac{1}{\pi
m_{i}\omega_{i}}\right)^{1/4}\exp{\left(-\frac{p_{i}^{2}}{2m_{i}\omega_{i}}
+\emph{i}\,\sqrt{\frac{2}{m_{i}\omega_{i}}}\,p_{i}a_{i}^{\dag}
+\frac{1}{2}a_{i}^{\dag 2}\right)}|0>_{i},
\end{equation}
where $a_{i}^{\dag}$ (and $a_{i}$) are the ordinary bosonic
creation (and annihilation) operators
\begin{equation}
a_{i}=\frac{1}{2}\left(\sqrt{m_{i}\omega_{i}}\eta_{i}
+\emph{i}\,\frac{1}{\sqrt{m_{i}\omega_{i}}}p_{i}\right),~~~~
a_{i}^{\dag}=\frac{1}{2}\left(\sqrt{m_{i}\omega_{i}}\eta_{i}
-\emph{i}\,\frac{1}{\sqrt{m_{i}\omega_{i}}}p_{i} \right).
\end{equation}
It is not difficult to see that $V$ transforms
$Vp_{i}V^{\dag}=b_{ij}p_{j}$ and $V
\eta_{i}V^{\dag}=a_{ji}\eta_{j}$, because
\begin{equation}
Vp_{i}V^{\dag}=det\,A\,\int d\vec{p}\,d\vec{p}\prime|A\vec{p}>\,\,
<\vec{p}|p_{i}|\vec{p}\prime>\,<A\vec{p}\prime| =\int
d\vec{\tilde{p}}\,b_{ij}\tilde{p}_{j}\,|\vec{\tilde{p}}>\,
<\vec{\tilde{p}}|=b_{ij}p_{j},
\end{equation}
and
\begin{equation}
V\eta_{i}V^{\dag}=det\,A\,\int d\vec{p}\,d\vec{p}\prime\,
|A\vec{p}>\,\emph{i}\frac{\partial}{\partial p_{i}}\, \delta
(\vec{p}-\vec{p}\prime )\,<A\vec{p}\prime|,
\end{equation}
furthermore, acting eq.(72) from the right-hand side on
$<\vec{\eta}|$ leads to
\begin{equation}
<\vec{\eta}|\,V\eta_{i}V^{\dag}=det\,A \int
d\vec{p}\,d\vec{p}\prime\, \left(-\emph{i}\frac{\partial}{\partial
p_{i}}\, \exp{(\emph{i}\,a_{jk}p_{k}\eta_{j})}\right)\,\delta
(\vec{p}-\vec{p}\prime)\, <A\vec{p}\prime
|=<\vec{\eta}|a_{ji}\,\eta_{j},
\end{equation}
which means that $V \eta_{i}V^{\dag}=a_{ji}\eta_{j}$.

\par
Now let us act the unitary transformation $V$ on the Hamiltonian
(66) and get
\begin{eqnarray}
VHV^{\dag}&=&\frac{1}{2m_{1}}(b_{11}p_{1}+b_{12}p_{2})^{2} +
\frac{1}{2m_{2}}(b_{21}p_{1}+b_{22}p_{2})^{2} -\alpha
(b_{11}p_{1}+b_{12}p_{2})\,(b_{21}p_{1}+b_{22}p_{2})\nonumber\\
&\,&+\frac{m_{1}\omega_{1}^{2}}{2}(a_{11}\eta_{1}+a_{21}\eta_{2})^{2}
+\frac{m_{2}\omega_{2}^{2}}{2}(a_{12}\eta_{1}+a_{22}\eta_{2})^{2}\nonumber\\
&\,&-\beta
(a_{11}\eta_{1}+a_{21}\eta_{2})\,(a_{12}\eta_{1}+a_{22}\eta_{2}).
\end{eqnarray}
Then in order to annihilate the coupling terms in eq.(74), we set
\begin{eqnarray}
\frac{1}{m_{1}}a_{22}a_{12}+\frac{1}{m_{2}}a_{21}a_{22}+ \alpha
(a_{11}a_{22}+a_{12}a_{21})&=&0,\nonumber\\
m_{1}\omega_{1}^{2}\,a_{11}a_{21}+m_{2}\omega_{2}^{2}\,a_{12}a_{22}
-\beta (a_{11}a_{22}+a_{12}a_{21})&=&0.
\end{eqnarray}
From eq.(75) we have
\begin{equation}
a_{12}=\frac{\Omega m_{1}}{2\alpha(\beta+\alpha
m_{1}m_{2}\omega_{2}^{2})}\,a_{11},~~~~~~~~ a_{21}=-\frac{\Omega
m_{2}}{2\alpha(\beta+\alpha m_{1}m_{2}\omega_{1}^{2})}\,a_{22},
\end{equation}
where
\begin{equation}
\Omega =\alpha (\omega_{1}^{2}-\omega_{2}^{2})+
\sqrt{\alpha^{2}(\omega_{1}^{2}-\omega_{2}^{2})^{2}+ 4\alpha^{2}
(\frac{\beta}{m_{1}}+\alpha
m_{2}\omega_{2}^{2})\,(\frac{\beta}{m_{2}}+\alpha
m_{1}\omega_{1}^{2}) }.
\end{equation}
Thus eq.(74) can be written as
\begin{eqnarray}
H_{d}&=&\frac{a_{22}^{2}}{2m_{1}(det\,A)^{2}}
\left(1+\frac{m_{1}m_{1}\Omega}{\beta+\alpha
m_{1}m_{2}\omega_{1}^{2}}+ \frac{m_{1}m_{2}\Omega^{2}}{4\alpha^{2}
(\beta+\alpha
m_{1}m_{2}\omega_{1}^{2})^{2}}\right)p_{1}^{2}\nonumber\\
&\,& + \frac{a_{11}^{2}}{2m_{2}(det\,A)^{2}}
\left(1-\frac{m_{1}m_{1}\Omega}{\beta+\alpha
m_{1}m_{2}\omega_{2}^{2}}+ \frac{m_{1}m_{2}\Omega^{2}}{4\alpha^{2}
(\beta+\alpha m_{1}m_{2}\omega_{2}^{2})^{2}}\right)p_{2}^{2}
\nonumber\\
&\,&+\frac{a_{11}^{2}m_{1}\omega_{1}^{2}}{2}\left(1-\frac{\beta
\Omega}{\alpha \omega_{1}^{2}(\beta+\alpha
m_{1}m_{2}\omega_{2}^{2})}+\frac{m_{1}m_{2}\omega_{2}^{2}
\Omega^{2}}{4\alpha^{2}\omega_{1}^{2}(\beta+\alpha
m_{1}m_{2}\omega_{2}^{2})^{2}}\right)\eta_{1}^{2}\nonumber\\
&\,&+\frac{a_{22}^{2}m_{2}\omega_{2}^{2}}{2}\left(1+\frac{\beta
\Omega}{\alpha \omega_{2}^{2}(\beta+\alpha
m_{1}m_{2}\omega_{1}^{2})}+\frac{m_{1}m_{2}\omega_{1}^{2}
\Omega^{2}}{4\alpha^{2}\omega_{2}^{2}(\beta+\alpha
m_{1}m_{2}\omega_{1}^{2})^{2}}\right)\eta_{2}^{2}.
\end{eqnarray}
Since
\begin{equation}
det\,A=a_{11}a_{22}-a_{12}a_{21}=a_{11}a_{22}\left(
1+\frac{m_{1}m_{2}\Omega^{2}}{4\alpha^{2}(\beta+\alpha
m_{1}m_{2}\omega_{1}^{2})(\beta+\alpha m_{1}m_{2}\omega_{2}^{2})}
\right),
\end{equation}
if we use the following notations
\begin{eqnarray}
T_{1}&=&1+\frac{m_{1}m_{2}\Omega^{2}}{4\alpha^{2}(\beta+\alpha
m_{1}m_{2}\omega_{1}^{2})(\beta+\alpha
m_{1}m_{2}\omega_{2}^{2})},\nonumber\\
T_{2}&=&1+\frac{m_{1}m_{1}\Omega}{\beta+\alpha
m_{1}m_{2}\omega_{1}^{2}}+ \frac{m_{1}m_{2}\Omega^{2}}{4\alpha^{2}
(\beta+\alpha m_{1}m_{2}\omega_{1}^{2})^{2}},\nonumber\\
T_{3}&=&1-\frac{\beta \Omega}{\alpha \omega_{1}^{2}(\beta+\alpha
m_{1}m_{2}\omega_{2}^{2})}+\frac{m_{1}m_{2}\omega_{2}^{2}
\Omega^{2}}{4\alpha^{2}\omega_{1}^{2}(\beta+\alpha
m_{1}m_{2}\omega_{2}^{2})^{2}},\nonumber\\
T_{4}&=&1-\frac{m_{1}m_{1}\Omega}{\beta+\alpha
m_{1}m_{2}\omega_{2}^{2}}+ \frac{m_{1}m_{2}\Omega^{2}}{4\alpha^{2}
(\beta+\alpha m_{1}m_{2}\omega_{2}^{2})^{2}},\nonumber\\
T_{5}&=&1+\frac{\beta \Omega}{\alpha \omega_{2}^{2}(\beta+\alpha
m_{1}m_{2}\omega_{1}^{2})}+\frac{m_{1}m_{2}\omega_{1}^{2}
\Omega^{2}}{4\alpha^{2}\omega_{2}^{2}(\beta+\alpha
m_{1}m_{2}\omega_{1}^{2})^{2}},
\end{eqnarray}
and in eq.(78), let the coefficients of the terms
$p_{1}^{2}/2m_{1}$ and $p_{2}^{2}/2m_{2}$ be equal to the
coefficients of the terms $m_{1}\omega_{1}^{2}\eta_{1}^{2}/2$ and
$m_{2}\omega_{2}^{2}\eta_{2}^{2}/2$ respectively, and denote them
as $\Lambda_{1}$ and $\Lambda_{2}$, we have
\begin{eqnarray}
&&a_{11}= T_{1}^{-1/2}T_{2}^{1/4}T_{3}^{-1/4},~~~~ a_{22}=
T_{1}^{-1/2}T_{4}^{1/4}T_{5}^{-1/4},\nonumber\\
&&\Lambda_{1}=T_{1}^{-1}T_{2}^{1/2}T_{3}^{1/2},~~~~
\Lambda_{2}=T_{1}^{-1}T_{4}^{1/2}T_{5}^{1/2}.
\end{eqnarray}
Thus we diagonalize the Hamiltonian (64) and obtain
\begin{equation}
H_{d}=VHV^{\dag}=\Lambda_{1}\omega_{1}\left(a_{1}^{\dag}a_{1}+\frac{1}{2}
\right) +\Lambda_{2}\omega_{2}\left(a_{2}^{\dag}a_{2}+\frac{1}{2}
\right),
\end{equation}
which gives the energy spectrum of the two-dimensional harmonic
oscillator (62) on the noncommutative plane with both the kinetic
and the elastic couplings
\begin{equation}
E_{n,m}=\Lambda_{1}\omega_{1}\left(n+\frac{1}{2}\right)+
\Lambda_{2}\omega_{2}\left(m+\frac{1}{2}\right).
\end{equation}
This result, to our knowledge, has not been reported in the
literature so far. In some special case, however, it reduces to
well-known relevant results. For example, when the coupling
constants $\kappa$ and $\lambda$ both vanish, the Hamiltonian (62)
describes a two-dimensional harmonic oscillator without any
coupling on the noncommutative plane. Eq.(83) reduces to
\begin{equation}
E_{n,m}=\sqrt{1+\frac{\theta^{2}}{4}} \,(n+m+1),
\end{equation}
which was derived by many authors in other methods. For instance,
eq.(84) coincides with \cite{s8}.

\section{Summary and discussion}
In order to develop representation and transformation theory so
that one can solve more dynamic problems for NCQM, in this work we
introduce new representations on the noncommutative space which
may be named the entangled state representations, because the
sate-vectors of these representations are common eigenstates of
the difference (or the sum) of two different coordinate-component
operators and the sum (or the difference) of two relevant momentum
operators, and display some entanglements of different components
on the noncommutative space. Since these state-vectors are
orthonormal and satisfy the completeness relation, they form
representations to formulate the NCQM. In this work we find out
explicit unitary operator which can transfers the entangled state
representation $|\eta>$ into the so-called "quasi-coordinate"
representation $|x,y>$ used in the papers on NCQM. Similar unitary
operator between the $|\xi>$ representation and the $|x,y>$
representation can be got also. To show the potential applications
of new entangled representations, we solve exactly a
two-dimensional harmonic oscillator with both the kinetic and the
elastic couplings on the noncommutative plane. This example shows
that some dynamic problems of NCQM may be easier solved in the
entangled state representations.

\par
It is also interesting to generalize the entangled state
representations to describe two particles moving on the
noncommutative space. Work on this direction will be presented in
a separate paper.


\begin{thebibliography}{10}
\bibitem{s1} P. A. M. Dirac, \emph{The Principles of Quantum
Mechanics}, Oxford Clarendon Press, 1930.
\bibitem{s2} A. Connes, M. R. Douglas and A. Schwartz,
\emph{Noncommutative geometry and matrix theory: Compactification
on tori}, \emph{JHEP} 9802 (1998) 003; N. Seiberg and E. Witten,
\emph{String theory and noncommutative geometry}, \emph{JHEP} 9909
(1999) 032; P-M. Ho and H-C. Kao, \emph{Noncommutative quantum
mechanics from noncommutative quantum field theory},
\emph{hep-th/0110191}; D. Kochan and M. Demetrian, \emph{Quantum
mechanics on noncommutative plane}, \emph{hep-th/0102050}.
\bibitem{s3} J. E. Moyal, \emph{Quantum mechanics as a statistical
theory}, \emph{Proc. Cambridge Phil. Soc.} 45 (1949) 99.
\bibitem{s4} A. Einstein, B. Podolsky and N. Rosen, \emph{Can quantum
mechanical description of Physical reality be considered
complete?} \emph{Phys. Rev.} \textbf{47} (1935) 777.
\bibitem{s5} H. Fan and J. R. Klauder, \emph{Eigenvectors of two particles'
relative position and total momentum} \emph{Phys. Rev.}
\textbf{A49} (1994) 704.
\bibitem{s6} H. Fan, H. R. Zaidi and J. R. Klauder, \emph{New approach for
calculating the normally ordered form of squeeze operators}
\emph{Phys. Rev.}
\textbf{D35} (1987) 1831.
\bibitem{s7} B. Muthukumar and P. Mitra, \emph{Noncommutative oscillators
and the commutative limit}, \emph{Phys. Rev. D}\textbf{66} (2002)
027701; V. P. Nair and A. P. Polychronakos, \emph{Quantum
mechanics on the noncommutative plane and sphere},
\emph{hep-th/0011172}.
\bibitem{s8} A. Jellal, \emph{Orbital magnetism of a two-dimensional
noncommutative confined system}, \emph{J. Phys. A: Math. Gen.}
\textbf{34} (2001) 10159.
\end{thebibliography}
\end{document}